\documentclass[%
 reprint,
 amsmath,amssymb,
 aps,nofootinbib, longbibliography, floatfix
]{revtex4-1}
\usepackage{amsmath, amsthm, amssymb, amsfonts}
\usepackage{bm}
\PassOptionsToPackage{linktocpage}{hyperref}
\usepackage[hyperindex,breaklinks]{hyperref}
\usepackage{enumitem}
\usepackage{slashed}

\renewcommand{\theta}{\vartheta}

\usepackage{array}
\usepackage{mathtools}
\usepackage{xcolor}

\usepackage{etoolbox}
\smallskipamount
\begin{document}

\title{${SO(10)}$: a Case for Hadron Colliders}

\author{Ance Preda$^{1}$, Goran Senjanovi\'c$^{1,2}$ and Michael Zantedeschi$^{1,3}$}
\affiliation{%
$^1$
Arnold Sommerfeld Center, Ludwig-Maximilians University, Munich, Germany
}%
\affiliation{%
$^2$
International Centre for Theoretical Physics, Trieste, Italy
}%
\affiliation{%
$^3$
Max-Planck-Institute for Physics, Munich, Germany
}%

\begin{abstract}
We study the mass scales in the $SO(10)$ grand unified theory based on the following minimal Higgs representation content: adjoint $45_{\rm H}$, spinor $16_{\rm H}$ and complex vector $10_{\rm H}$, with higher dimensional operators on top of renormalizable interactions. Assuming no judicial cancellations in proton decay amplitudes, the consistency of the theory requires scalar weak triplet, scalar doublet leptoquark and scalar gluon octet to lie roughly below $ 10\, \rm TeV$ energy and potentially accessible even at the LHC.
Under the same assumption, these signatures are intimately connected with an expectation of proton lifetime below $10^{35} {\rm yr}$, to be probed in the new generation of proton decay experiments. 
\end{abstract}

\maketitle

\section{Introduction}
Grand Unification has been one of the main candidate theories beyond the Standard Model (SM). It provides a rationale for charge quantization in nature and leads to the prediction of proton decay and the existence of magnetic 
monopoles~\cite{Polyakov:1974ek,tHooft:1974kcl}. 
With time, the original enthusiasm for the unification of the SM vaned since the gauge couplings do not unify, as it originally seemed~\cite{Georgi:1974yf} - $\alpha_1$ meets $\alpha_2$ too early. Since the SM augmented with low energy supersymmetry predicted gauge coupling unification~\cite{Dimopoulos:1981yj,Ibanez:1981yh,Einhorn:1981sx,Marciano:1981un} at energy on the order of $10^{16} {\rm GeV}$, the field turned to supersymmetric grand unification.
  
However, grand unified theories necessarily contain new physical states whose contribution to the evolution of gauge couplings may change the SM predictions. It is thus important to reassess the original, non-supersymmetric, grand unification.
 
 The minimal $SU(5)$ gauge symmetry~\cite{Georgi:1974sy} is a logical starting point, but just as the SM, it fails to unify gauge couplings and is tailor made for massless neutrinos.  While there are minimal non-renormalisable extensions that can cure both problems~\cite{Dorsner:2005fq,Bajc:2006ia}  - and moreover~\cite{Bajc:2006ia} predicts new light states at today's energies - 
 it is natural to 
 turn to the $SO(10)$ theory, which from the outset implied non-vanishing neutrino mass, and moreover unifies a generation of fermions, otherwise fragmented in $SU(5)$. 
 Furthermore, it has been text-book wisdom for decades that the gauge couplings unify naturally in $SO(10)$ through an intermediate scale $M_{\rm I}$ and the desert between $M_{\rm W}$ and $M_{\rm I}$
 
 In this Letter, we revisit the $SO(10)$ model with small representations, 
which allows to treat unification constraints without any assumption on the particle mass spectra. Surprisingly, 
we find that the desert picture fails against all prejudice. Instead, we find a nearby oasis with the following scalar particles at today's energies: scalar gluon octet, scalar analogs of $W$ and $Z$ and scalar leptoquark doublet. Moreover, these predictions are tied to a likely observability of proton decay in the planned experiments.\\

\section{$SO(10)$ Theory}
The $SO(10)$ symmetry group \cite{Georgi:1974my,Fritzsch:1974nn}, 
by unifying a generation of fermions in a spinor representation $16_{\rm F}$, predicts a Right-Handed (RH) neutrino and a small neutrino mass through the seesaw mechanism~\cite{Minkowski:1977sc,Mohapatra:1979ia,Glashow:1979nm,GellMann:1980vs}. 
The gauge coupling unification is naturally achieved~\cite{Shafi:1979qb,delAguila:1980qag,Rizzo:1981su} through an intermediate scale in the form of the Left-Right(LR) ~\cite{Mohapatra:1974gc,Senjanovic:1975rk,Senjanovic:1978ev} or the Pati-Salam quark-lepton (QL)~\cite{Pati:1974yy} symmetries.

The minimal version of the theory (with small representations) contains three $16_{\rm F}$ spinors (fermion generations), augmented with the following Higgs scalars
\begin{equation}\label{higgs}
45_{\rm H}; \,\,\,\,\,  16_{\rm H}; \,\,\,\,\,  10_{\rm H}\,,
\end{equation}
where the representation content is specified in obvious notation. The $45_{\rm H}$ field is an adjoint, used for the GUT symmetry breaking down to an intermediate scale which is then broken by the vacuum expectation value of the spinor $16_{\rm H}$ to the SM gauge symmetry. Finally, a complex $10_{\rm H}$ vector is used to complete the breaking down to charge and color gauge invariance, in the usual manner. 

The minimal Higgs sector would employ a real $10_{\rm H}$, but then we would have a single Yukawa coupling and a single vacuum expectation value, predicting all fermion masses being equal (in particular incurable top-bottom mass equality) - thus the need to complexify $10_{\rm H}$. 

Even the complex $10_{\rm H}$ cannot suffice at the renormalizable $d=4$ level, since it predicts equal down quark and charged lepton masses. The way out is through the addition of higher-dimensional operators, whose contribution is suppressed by $\langle 45_{\rm H} \rangle /\Lambda$, $\Lambda$ being the scale where gravity becomes strong (or the scale of some new physics between that scale and the GUT one). In what follows we take $\Lambda \gtrsim 10 M_{\rm GUT}$, in order for this expansion to be perturbatively valid.

Notice though that in the case of the third generation, the dominant contribution comes from the tree level Yukawa coupling in order to guarantee the large top quark mass. This implies an important relation for the third generation neutrino Dirac mass term 
\begin{equation}\label{thirddirac}
m_{\rm D3} = m_{\rm t}\,.
\end{equation}
This relation plays a crucial role in the rest of the paper. 

It is well known that at the tree level $\langle 45_{\rm H} \rangle $ keeps $ {SU(5) \times U(1)}$ symmetry unbroken. Since $\langle 16_{\rm H} \rangle $ lies in the $ {SU(5)}$ singlet direction, the theory appears to be unrealistic. 
However, when the effective Coleman-Weinberg potential~\cite{Coleman:1973jx} is taken into account~\cite{Bertolini:2009es}, besides the $ {SU(5) \times U(1)}$ case, $\langle 45_{\rm H} \rangle$ breaks $SO(10)$ to an intermediate symmetry based on 
$SU(2)_{\rm L} \times SU(2)_{\rm R} \times U(1)_{\rm B-L} \times SU(3)_{\rm C}$ (LR) or $ SU(2)_{\rm L} \times U(1)_{\rm R} \times SU(4)_{\rm C}$ (QL) gauge groups, and then $\langle 16_{\rm H} \rangle $ completes the breaking down to the SM as required. It turns out, though, that the scalar masses end up being constrained.

On the other hand, it is straightforward to show (the details are left for a longer paper, now in preparation \cite{PSZ}) that the inclusion of higher-dimensional operators in the scalar potential
 allows for the realistic symmetry breaking as above - however without restrictions on the mass spectrum. We will stick to this in what follows in order to be as general as possible and claim predictions independent of the parameter space. 
 
 Before we plunge into details, a comment is in order. From the failure of the minimal $ {SU(5)}$ theory to successfully unify gauge couplings, one expects single step breaking of $ {SO(10)}$ or the ${SU(5)}$ intermediate symmetry  to fail similarly - and indeed both do, as we have verified. In short, one needs LR or QL intermediate symmetry for unification to work, and hereafter we shall stick to them.

The vacuum expectation values of the $45_{\rm H}$ can thus be written in the canonical form 
 \begin{eqnarray}\label{45vev}
&\langle 45_{\rm H} \rangle^{\rm LR}& = v_{\rm GUT}\, i\,\sigma_2\, \otimes {\rm diag} (1, 1, 1, 0, 0)  \nonumber  \\
&\langle 45_{\rm H} \rangle^{\rm QL} &= v_{\rm GUT}\, i\,\sigma_2\, \otimes {\rm diag} (0, 0, 0, 1, 1)\,,\label{vev}
\end{eqnarray}
for the LR and QL preserving cases, respectively. The potentially possible case where both vevs above do not vanish, would leave 
$SU(2)_{\rm L} \times U(1)_{\rm R} \times U(1)_{\rm B-L} \times SU(3)_{\rm C}$ symmetry, which from the point of view of running is equivalent to the SM symmetry and thus also ruled out.

Once $\langle 16_{\rm H} \rangle = M_{\rm I} $ gets turned on, through the interaction $16_{\rm H} \,45_{\rm H} \,16^*_{\rm H}$, the zeroes in \eqref{vev} get corrected by $M_{\rm I}/M_{\rm GUT}$.
As explained above, unification constraints require $M_{\rm GUT} \gg M_{\rm I}$.

Independently of the choice of the intermediate symmetry, neutrino mass plays an essential role in constraining the scales of symmetry breaking. The argument goes as follows. 

The RH neutrino (N) mass originates from the $d=5$ operator 
$ {16_{\rm F} 16_{\rm F} 16_{\rm H}^* 16_{\rm H}^*}/\Lambda$ which gives
\begin{equation}\label{nuReffect}
m_{\rm N} \simeq \frac {M_{\rm I}^2}{\Lambda}\,,
\end{equation}
where $M_{\rm I}$ is the intermediate-mass scale corresponding either to LR or QL symmetry.
There is also a well-known two-loop diagram~\cite{Witten:1979nr}, which amounts to \\$m_{\rm N} \simeq(\alpha / \pi)^2 M_{\rm I}^2 / M_{\rm GUT} $. 

Clearly, the latter contribution is necessarily smaller since $\Lambda$ must lie below the Planck scale. It has been argued convincingly 
that $\Lambda \lesssim M_{\rm Pl} / \sqrt {N}$~\cite{Dvali:2007hz,Dvali:2007wp}, where $N$ is the number of degrees of freedom of the theory in question, at least of order $10^2$, strengthening the case for a higher dimensional source for RH neutrino mass.

Using then \eqref{nuReffect} and \eqref{thirddirac}, one gets for the neutrino mass from the seesaw mechanism
\begin{eqnarray}\label{nueffect}
m_{\rm \nu} \simeq  \frac {(m_{\rm 3 D})^2} {m_{\rm N}}  \simeq \frac {m_{\rm t}^2 \Lambda} {M_{\rm I}^2} \simeq \nonumber\qquad\qquad\qquad\qquad\qquad\\
	\simeq {\rm eV}\left(\frac{m_{\rm t}} {100 \rm {GeV}} \frac{6\cdot 10^{14}\rm Gev}{M_{\rm I}}\right)^2  \left(\frac{\Lambda}{4 \cdot 10^{16}{\rm GeV}} \right),
\end{eqnarray}
where we normalize the scales in question by the most suitable choice, see the Table below. Eq.~\eqref{nueffect} is valid at $M_{\rm GUT}$. The running of $m_\nu$ to high energies\footnote{We thank the Referee for reminding us of this correction.} \cite{Ray:2010rz} approximately doubles its value. Notice that $m_{\rm t} \simeq 100 {\rm GeV}$ at $M_{\rm GUT}$, see e.g. Table 3 in~\cite{Babu:2016bmy}.

Combining this with the direct upper limit on neutrino mass from KATRIN experiment~\cite{Aker:2022ijt} $m_\nu < 0.8 {\rm eV}$ leads to $M_{\rm I}\gtrsim 4\times 10^{14}\rm GeV$, where we allow a relative uncertainty of a factor $2$ in \eqref{nuReffect}.
There is also an indirect 
GERDA limit 
$m_\nu \lesssim 0.2\, {\rm eV}$~\cite{GERDA:2020xhi} from neutrinoless double beta decay, relevant for Majorana neutrinos, but due to the possible mixing angles suppression less relevant. We will stick here to $m_\nu < 0.8 {\rm eV}$ - lowering it only strengthens our results. 

The smallness of neutrino mass sends a clear message: the intermediate scale in this $SO(10)$ theory must be huge, close to the GUT scale. We are thus in a situation very similar to the minimal $SU(5)$ theory - and the learned reader~\cite{Bajc:2006ia} can guess that there ought to be some light states at today's energies in order for the model to unify.  

At the same time, the cutoff $\Lambda$ should be as small as possible. Since, on the other hand, $\Lambda \gtrsim 10\, M_{\rm GUT}$, this, in turn, implies that the GUT scale should be as low as possible, making a case for the potential discovery of proton decay.  The issue however is the connection between the proton lifetime and the unification scale, which depends on whether one is willing to accept judicial cancellations~\cite{Nandi:1982ew,Dorsner:2004xa} in the proton decay amplitudes. In the rest of this work, we follow both the conventional and established approach of shying away from such a possibility. Therefore, from the  bound on proton lifetime $\tau_{\rm p} \gtrsim 10^{34} \,{\rm yr}$~\cite{Super-Kamiokande:2020wjk}, one gets $M_{\rm GUT} \gtrsim 4 \cdot 10^{15} {\rm GeV}$. \\

\section{Unification Constraints}
As we said above, in the SM $\alpha_1$ and $\alpha_2$ unify too early. An intermediate scale $M_{\rm I}$ with $U(1)$ embedded into a non-abelian symmetry helps since it slows down the rise of $\alpha_1$. One would expect $M_{\rm I}$ much below $M_{\rm GUT}$ - but as we shall see, this simply implies too large neutrino mass. 

Let us first elaborate on the particle content of the theory. Besides the three generations of fermions and the SM gauge bosons, we also have the following scalar particles with non-vanishing SM quantum numbers (thus present in renormalization group equations). 

From $45_{\rm H}$ and $16_{\rm H}$: scalar gluons with mass $m_8$ and scalar $W,Z$ states with mass $m_3$, scalar up quark with mass $m_{\rm sup}$, 
 scalar lepton doublet with mass $m_{\rm sl}$, scalar quark doublet with mass $m_{\rm sq}$, scalar down quark with mass $m_{\rm sdown}$ and scalar electron with mass $m_{\rm sel}$. The reader should not be confused with our language borrowed from supersymmetry - it is just a shorthand to particle properties. 

From $10_{\rm H}$: two Higgs doublets (including the SM one), and their color triplet partners that mediate proton decay and are forced to lie close to the GUT scale. 

Our main result stems from the following unification condition 
\begin{eqnarray}
\frac {M_{\rm GUT}} {M_{\rm Z}} \simeq \exp \left[{\frac{\pi}{10}\left(5\alpha_{1}^{-1}-3\alpha_{2}^{-1}-2\alpha_{3}^{-1}\right)}\right]\nonumber
\\
\cdot \left[\left(\frac{M_{\rm Z}}{M_{\rm I}}\right)^{22}\left(\frac{M_{\rm Z}^{2} m_{\rm sel} m_{\rm sup}}{m_{3}m_{8}m_{\rm sq}^{2}}\right)\right]^{\frac{1}{20}} \,,
\label{mainconstraint}
\end{eqnarray}
 where couplings are evaluated at the scale $M_{\rm Z}$. It is based on the one-loop renormalization group flow valid for both the LR and QL breaking patterns. Higher-dimensional effects in the gauge coupling sector are ignored for the moment. 

As we said above, the crux is to have $M_{\rm GUT}$ large enough for the sake of the proton's longevity. Before LEP, when $\alpha_1$ was thought to be smaller, it seemed that the exponential pre-factor term could do the job even without any intermediate symmetry and with all new states superheavy, but today we know that it does not work. An obvious way out is small $M_{\rm I} \lesssim 10^{13} \rm GeV$ as usually assumed, but as \eqref{nueffect} shows, neutrino mass considerations force a large $M_{\rm I} \gtrsim 4\times10^{14} {\rm GeV}$. Thus the burden is then on the scalar states to save the theory.

In other words, $M_{\rm Z}^{2} m_{\rm sel}m_{\rm sup}/  m_{3}m_{8}m_{\rm sq}^{2} $ has to be as large as possible.
 Clearly, the color octet, weak triplet and a scalar quark doublet field contribute in the right direction if light, while the scalar electron and scalar up-quark states have the opposite effect. A numerical estimate shows that former states ought to lie close to the electroweak scale, and the latter ones are forced to be heavy, close to the GUT scale. The other fields simply decouple from this particular combination of couplings. 
 
To support our results, a two-loops analysis was performed. We spare the reader of the computational tedium here - the numerical details of this complete analysis will be presented in a longer paper, now in preparation~\cite{PSZ}. 
For the present discussion it should be noted that unification can be achieved only for $M_{\rm I}\lesssim 6 \times 10^{14}\rm GeV$. Simply, the effect of an even higher $M_{\rm I}$ cannot be traded for light scalar states.
It follows then that the intermediate scale is effectively fixed due to its lower bound coming from neutrino mass expression \eqref{nueffect}, $M_{\rm I} \simeq 5\times10^{14} {\rm GeV}$.

Significant corrections to the running come from higher-dimensional operators. In fact, the couplings do not have to unify at the GUT scale due to the possible presence of higher dimensional effective kinetic energy terms~\cite{Shafi:1983gz} 
\begin{equation}\label{kinetic}
{\rm Tr} F_{\mu \nu} F^{\mu \nu}   \frac {\langle 45_{\rm H} \rangle^2} {\Lambda^2} \,.
\end{equation}
It is easy to see that the liner term in $\langle 45_{\rm H} \rangle$ vanishes due to the asymmetry of the adjoint representation. Using\eqref{vev} one gets 
($\epsilon = M_{\rm GUT}/\Lambda$)
\begin{equation}\label{shiftLR}
\delta \alpha_1^{\rm LR} =\delta \alpha_1^{\rm QL} = \epsilon^2 \,;\quad\delta \alpha_3^{\rm LR} = \frac {5}{2} \epsilon^2 \,;\quad\delta \alpha_2^{\rm QL} = \frac {5}{3} \epsilon^2\,, 
\end{equation}
(and zero otherwise), for the LR and QL cases, respectively, with . The relative factors 5/2 and 5/3 come from the normalization of the $U(1)$ generator. Interestingly, for the QL case, these effects can help raise both $M_{\rm I}$, needed for neutrino mass and $M_{\rm GUT}$ in order to ensure proton stability. It should be noted that corrections to the zeroes in \eqref{45vev} - of the order $(M_{\rm I}/M_{\rm GUT})^2$  - to this small contribution can be safely ignored.\\

\section{Predictions} 
The LR symmetry breaking pattern ends up being ruled out since the intermediate scale turns out to be not higher than about $M_{\rm I}\lesssim10^{14}\rm GeV$, too small to be compatible with neutrino mass in \eqref{nueffect}. 
On the other hand, the QL breaking pattern provides a viable scenario with $M_{\rm I}\simeq 10^{15}{\rm GeV}$, as required by the smallness of neutrino mass. This reduction of the parameter space makes the theory even more predictive. 

There is more to it. The QL case does not correct $m_{\rm d} = m_{\rm e}$ relations, however, $\langle 45_{\rm H} \rangle$ in \eqref{45vev} gets corrections from $\langle 16_{\rm H} \rangle$ which breaks the QL symmetry. In other words, the relation $m_{\rm b} = m_{\rm\tau}$ gets corrected by $M_{\rm I}/\Lambda$.
A possible operator accounting for the required correction is $16_{\rm F}^2 45_{\rm H} 10_{\rm H}$. Notice that $45_{\rm H}10_{\rm H}\supset 120$ which breaks QL symmetry when going through the $45_{\rm H}$ induced VEV\footnote{Another coupling of similar magnitude is $16_{\rm F}^2 16_{\rm H}^2$ where the Higgs doublet VEV in $16_{\rm H}$ is induced upon electroweak symmetry breaking. Effectively this also generates a $120$ correction to Yukawa couplings, therefore not affecting our discussion.}. 
Such contribution is, however, antisymmetric\footnote{We thank the referee for noticing this.} which still keeps the down quarks and charged lepton masses over-constrained, and no realistic spectrum emerges.

However, further corrections come from $d=6$ operators such as $16_{\rm F}^2 45_{\rm H}^2 10_{\rm H}\supset 126$ (and thus symmetric in generation space), easily seen to be sufficiently large.
 This makes the QL breaking corrections to the Yukawa couplings effectively arbitrary, and ensures a viable low-energy spectrum.

This provides an independent argument for having the QL intermediate scale close to the GUT one, and at the same time, $\Lambda \simeq 10\, M_{\rm GUT}$. This happens precisely because the unification constraints eliminate the alternative LR scenario, in which case QL symmetry would be broken already at the GUT scale, and {\textit per se} $M_{\rm I}$ could be much lower. In other words, neutrino mass consideration was necessary to establish the large $M_{\rm I}$ argument, which was essential in establishing new light states.

\begin{table}[]
\centering
\begin{tabular}{ p{3.2cm}p{1.8cm}p{2.7cm}  }
 \hline
 \multicolumn{3}{c}{Spectrum} \\
 \hline
 Particle&$3_{\rm C}\,2_{\rm L}\,1_{\rm Y} $ &Mass range\\
 \hline
 scalar quark doublet  (leptoquark) &$(3,2,1/6)$		&$\lesssim 10\rm{TeV}$ \\
 weak triplet &$(1,3,0)$	&$\lesssim 10\rm{TeV}$\\
 color octet &$(8,1,0)$	&$\lesssim 10\rm{TeV}$\\
 scalar lepton doublet &$(1,2,-1/2)$	&$10^3{\rm{GeV}} - M_{\rm I}$\\
 second Higgs doublet &$(1,2,-1/2)$	&$10^3{\rm{GeV}} - M_{\rm GUT}$\\
 scalar down quark   &$(3,1,-1/3)$       &$10^{12}{\rm{GeV}} - M_{\rm GUT}$\\
 color triplet Higgs partners &$(\overline{3},1,1/3)$ &$10^{12}{\rm{GeV}} - M_{\rm GUT}$\\
 scalar up 	quark	&$(3,1,2/3)$	&$10^{14}{\rm{GeV}} - M_{\rm GUT}$\\
 scalar electron &$(1,1,1)$		&$10^{14}{\rm{GeV}} - M_{\rm GUT}$\\
 \hline
\end{tabular}
\caption{Mass spectrum for QL breaking pattern. $M_{\rm GUT}\simeq0.1 \Lambda\simeq 4\cdot 10^{15}{\rm GeV}$, $M_{\rm I}\simeq 5 \cdot 10^{14}\rm{GeV}$. The light particles have to satisfy the lower limits on their masses from the LHC.}
\label{table}
\end{table}
The resulting spectrum is reported in Table~\ref{table}. Necessity of unification requires the scalar weak triplet, the scalar quark weak doublet and the scalar color octet to have masses no larger than approximately $10\,{\rm TeV}$. The scalar quark  doublet is actually a leptoquark\footnote{Light leptoquarks have been long contemplated, see e.g.,~\cite{Senjanovic:1982ex,Dorsner:2005fq}.} 
 - we discuss this below in the short section on Phenomenology - an exciting possibility due to its exotic properties. 

 As expected from \eqref{mainconstraint}, the up quark-like and electron-like scalars need to be heavy and lie around $M_{\rm GUT}$ or at most two orders of magnitude below. Thus, if one of these states was to be observed in near future, it would serve to invalidate the theory. 
 
 On the other hand, the scalar lepton and second Higgs doublet impact the running weakly and their masses are basically arbitrary. In the Table, we show their masses varying from $10^3 {\rm GeV}$ up to the ${\rm GUT}$ scale (of course, they must lie above their experimental limits). 
 
 The scalar color triplets, the partners of Higgs doublets in $10_{\rm H}$ and the scalar down quark are taken to be be heavy, above $10^{12}{\rm GeV}$, because they mediate proton decay (they mix).  Since these states have small charges under the SM gauge group, they have negligible impact on the unification constraints and their masses are basically unconstrained. If one allows for cancellations in their Yukawa couplings so that proton decay would be suppressed, they could lie even at TeV energies~\cite{Dvali:1992hc,DSS}.
 
Finally, independently of the specific spectrum realization, there emerges an upper limit for the unification scale $ M_{\rm GUT} < 10^{16} \rm GeV$, which suggests, in turn, the upper limit on proton lifetime $\tau_{\rm p} \lesssim 10^{35}\rm yr $ - making the theory relevant also in view of next-generation proton decay experiments.
It is noteworthy that the unification constraints with high enough $M_{\rm I}$ require a low ${\rm GUT}$ scale, as does the independent consideration of neutrino mass from \eqref{nueffect}.

As we said before, these predictions follow from not assuming the flavour rotation of proton decay. However, if partial suppressions were allowed for, our predictions of light scalars would be weakened, and even lost for big enough cancellations in proton decay amplitudes. For example, for a unificaton scale $M_{\rm GUT}$ three times smaller than the one considered, our predictions would be completely lost. Namely, we found viable unification scenarios where all new particles states were heavier than $100\rm TeV$.
We should stress, though, that in such a case no grand unified model would be able to predict its particle spectra.\\

\section{Phenomenology} 
The light predicted states, the weak triplet, the color octet and the scalar quark doublet are easily produced through their gauge interactions. LHC has established lower limits on their masses on the order of TeV~\cite{Dorsner:2016wpm}.
While the color octet is clearly degenerate, the mass splittings of the scalar quark doublet are small compared to its mass.

What about their decays? The octet has  tiny couplings to fermions which arise from $d=5$ Yukawa terms and are of order $M_{\rm W}/\Lambda$, implying $\Gamma_8\simeq m_8 M_{\rm W}^2/(8\pi\, \Lambda^2 )\sim 10^{-28}\rm GeV$ for $m_8\simeq \mathcal{O}(\rm TeV)$.  

The scalar quark doublet has a number of possible decay channels and the situation becomes quite complex. For this reason we leave the discussion for a longer paper now in preparation~\cite{PSZ}. 

The weak triplet can decay in the same manner as the color octet through the $d=5$ interaction, but there is more to it.
 Through the effective terms $  10_{\rm H} 45_{\rm H} 10_{\rm H}$ and $10_{\rm H} 45_{\rm H}^2 10_{\rm H}$, one gets $\mu\,  \Phi^\dagger 3_{\rm H} \Phi$, where $\Phi$ stands for the SM Higgs doublet. This makes the triplet phenomenologically very interesting, since through the $\mu$-term its neutral component develops a vacuum expectation value
 \begin{equation}\label{3vev}
\langle 3_{\rm H} \rangle = v_3 \simeq  \mu \, \left (\frac {M_{\rm W}}  {m_3} \right)^2 \,.
\end{equation}

Notice that $v_3$ modifies the SM value of W-mass, while keeping the Z-boson mass unchanged. From the high-precision success of the SM, one gets $v_3 \lesssim \mathcal{O}({\rm GeV})$. This could be rather exciting in view of the recent CDF announcement~\cite{CDF:2022hxs} of a possible deviation from the SM $W-$ mass value. If true, the CDF result would imply 
$v_3\simeq 5 \rm GeV$. The fact that this triplet, added ad-hoc to the SM, can account for the CDF anomaly was pointed out in~\cite{Strumia:2022qkt,Perez:2022uil,Popov:2022ldh}.

In the context of grand unification, it turns out moreover that the low-energy effective theory of the triplet becomes rather predictive as pointed out~\cite{Senjanovic:2022zwy} in the context of a simple realistic $SU(5)$ grand unified theory~\cite{Bajc:2006ia}. In principle the  scalar quark doublet could impact the CDF result at the loop level through its mass splitting. However, as can be easily seen, it is simply too heavy and thus it effectively decouples. As long as the additional Higgs doublets are not fine-tuned to be light, the physics of the weak triplet is remarkably predictive: its decays into the SM particles are determined by its mass and the mixing $\theta= g\, v_3/ M_{\rm W}$, where $g= \sqrt{4\pi\, \alpha_2}$, induced by $v_3$, with the SM Higgs doublet~\cite{Senjanovic:2022zwy}. 

The triplet  contains a real CP-even field $H^0$ and the charged field $H^+$, with $m_{H^0}= m_{H^+}\equiv m_H$~\cite{Senjanovic:2022zwy}.
$H^0$ decays into  pairs of $W$, $Z$ and Higgs bosons $h^0$, while the charged component $H^+$ decays mainly into a $WZ$ and $Wh^0$, with the following  decay rates~\cite{Senjanovic:2022zwy}
\begin{equation}
\begin{split}
\label{eq:neutralbos}
     &\Gamma(H^0\rightarrow W^+W^-)\simeq 2\Gamma(H^0\rightarrow ZZ)\simeq2\Gamma(H^0\rightarrow h^0h^0)\\&\simeq \Gamma(H^+\rightarrow W^+h^0)
     \simeq\Gamma(H^+\rightarrow W^+Z) \simeq\theta^2\frac{g^2}{64 \pi} \frac{m_H^3 }{M_W^2}.
    \qquad
    \end{split}
\end{equation}

Notice that the gauge boson and Higgs final state rates are equal, since at high energies the SM symmetry breaking can be ignored. Moreover, since $H^0$ and $H^+$ are degenerate, all the decay rates are correlated.

The decay rate of $\Sigma $ into fermion-antifermion pairs is only relevant in the case of the third generation, with~\cite{Senjanovic:2022zwy}
\begin{equation}
\label{eq:fermiondecay}
  \Gamma(H^+\rightarrow t\overline b)\simeq  \Gamma(H^0\rightarrow t\overline t) \simeq \theta^2 \frac{3g^2}{{32}\pi}\frac{m_t^2m_H}{M_W^2}. 
\end{equation}
 Clearly, $H^0$ has to be heavy enough in order to decay into $ t \bar t$ final state. Moreover, if the triplet is heavier than the color octet and/or the scalar quark doublet, it can also decay into these final states. The above decay rates are however unaffected by these new channels.

\section{Summary and outlook}  
The $SO(10)$ grand unified theory is the minimal structure that unifies both the SM forces and a generation of quarks and leptons. 
We have revisited here the version of the theory based on the smallest possible Higgs representations: an adjoint, a spinor and a complex vector and studied the unification constraints.

We find that the consistency of the theory (assuming no cancellations in proton decay amplitudes) requires the existence of new physical states at energies accessible at the next hadron collider, if not already at the LHC - see Table~\ref{table}. 
One of these has quantum numbers of a  scalar quark doublet, others are new scalar colored octet and a scalar weak triplet. The last one is of particular phenomenological relevance since it generically develops a small VEV that modifies the W-mass. This can naturally account for the recent CDF result~\cite{CDF:2022hxs} in which case one would end up with a clear and predictive low-energy effective theory.

{\bf Acknowledgments} We are grateful to Borut Bajc for illuminating discussions and for pointing an error in the original version of the paper, and to Gia Dvali, Alejandra Melfo and Umberto Cotti for useful comments. G.S. thanks the Cookie Lab in Split for warm hospitality during the course of this work.

\bibliography{biblio}

\end{document}